\documentclass[twocolumn]{aastex631}

\usepackage{amsmath}
\usepackage{amssymb}

\makeatletter
\def\ff{\xdef\@thefnmark{}\@footnotetext}
\makeatother

\usepackage{color}
\usepackage{graphicx}

\usepackage{txfonts}

\usepackage[utf8]{inputenc}

\begin{document}

\title{Towards a realistic evaluation of transport coefficients in non-equilibrium space plasmas}

\author[0000-0002-1349-3663]{Edin Husidic}
\affiliation{Centre for mathematical Plasma Astrophysics, Department of Mathematics, KU Leuven, Celestijnenlaan 200B, 3001 Leuven, Belgium}
\affiliation{Department of Physics and Astronomy,
University of Turku, 20014 Turku, Finland}

\author[0000-0002-9530-1396]{Klaus Scherer}
\affiliation{Institut für Theoretische Physik, Lehrstuhl IV: Plasma-Astroteilchenphysik,  
Ruhr-Universität Bochum, D-44780 Bochum, Germany}
\affiliation{Research Department, Plasmas with Complex Interactions, 
Ruhr-Universität Bochum, D-44780 Bochum, Germany}

\author[0000-0002-8508-5466]{Marian Lazar}
\affiliation{Centre for mathematical Plasma Astrophysics, Department of Mathematics, KU Leuven, Celestijnenlaan 200B, 3001 Leuven, Belgium}
\affiliation{Institut für Theoretische Physik, Lehrstuhl IV: Plasma-Astroteilchenphysik,  
Ruhr-Universität Bochum, D-44780 Bochum, Germany}

\author[0000-0002-9151-5127]{Horst Fichtner}
\affiliation{Institut für Theoretische Physik, Lehrstuhl IV: Plasma-Astroteilchenphysik,  
Ruhr-Universität Bochum, D-44780 Bochum, Germany}
\affiliation{Research Department, Plasmas with Complex Interactions, 
Ruhr-Universität Bochum, D-44780 Bochum, Germany}

\author[0000-0002-1743-0651]{Stefaan Poedts}
\affiliation{Centre for mathematical Plasma Astrophysics, Department of Mathematics, KU Leuven, Celestijnenlaan 200B, 3001 Leuven, Belgium}
\affiliation{Institute of Physics, University of Maria Curie-Sk{\l}odowska, \\ 
Pl.\ M.\ Curie-Sk{\l}odowska 5, 20-031 Lublin, Poland}

\received{4 December 2021}
\revised{5 January 2022}
\accepted{11 January 2022}
\submitjournal{ApJ}

\begin{abstract}

{Recent studies have outlined the interest for the evaluation of transport coefficients in space plasmas, where the observed velocity
distributions of plasma particles are conditioned not only by the binary collisions, e.g., at low energies, but also by the energisation of particles
from their interaction with wave turbulence and fluctuations, generating the suprathermal Kappa-distributed populations.
This paper provides a first estimate of the main transport coefficients based on regularised Kappa distributions (RKDs), which, unlike standard Kappa distributions
(SKDs), enable macroscopic parameterisation without mathematical divergences or physical inconsistencies.
All transport coefficients derived here, i.e., the diffusion and mobility coefficients, electric conductivity, thermoelectric
coefficient and thermal conductivity, are finite and well defined for all values of $\kappa > 0$. 
Moreover, for low values
of $\kappa$ (i.e., below the SKD poles), the transport coefficients can be orders of magnitudes higher than the corresponding
Maxwellian limits, meaning that significant underestimations can be made if suprathermal electrons are ignored.}

\end{abstract}


\section{Introduction}\label{sec:intro}

Plasma is by far the most dominant state of perceivable matter in the universe. Due to the opportunity of in-situ measurements, the heliosphere is a plasma system of highest interest. The solar wind is emitted from the Sun as a continuous stream of electrons and protons, and fills the heliospheric bubble \citep{Marsch-2006, Verscharen-etal-2019}.
\ff{\!\!\!\!\!\!\!\!Corresponding author: Edin Husidic \\ E-mail: edin.husidic@kuleuven.be}
The high energy, as well as the dilute nature of space plasmas, point towards a reduced influence of binary collisions, so that there is no full relaxation to thermal equilibrium, characterised by Maxwellian distributions \citep{Pierrard-Lazar-2010}. Indeed, observations show nonthermal velocity distributions maintained for long periods \citep{Maksimovic-etal-2005, Zouganelis-etal-2005},
most probably due to the interaction of particles with fluctuations \citep{Vocks-Mann-2003,Marsch-2006,Vocks-etal-2008}. While wave-particle interactions occur at all heliocentric distances and their effects
become more relevant at larger distances ($>$ 1~AU), it is argued that Coulomb collisions between particles can still play a significant role toward lower distances from the Sun, i.e., lower than 1~AU \citep{Salem-etal-2003, Landi-etal-2010, Landi-etal-2012}. Indeed, solar wind models using a purely exospheric approach, that is, a collision-less model, fail to accurately reproduce the global
expansion of the solar wind observed in-situ, and are, at best, helpful approximations to gain insight into basic energetic processes (see the reviews by \cite{Marsch-1994} and \cite{Echim-etal-2011}). 
Thus, to realistically account for processes like dissipation, diffusion and viscosity, these models must also accommodate particle-particle collisions, or at least incorporate their effects, e.g., for low-energy populations (with high so-called collisional age) \citep{Bourouaine-etal-2011}. 

Space probes regularly observe nonthermal plasma particle velocity distributions with enhanced suprathermal tails, well described by the family of Kappa (or $\kappa$-power law) distributions
\citep{Pierrard-Lazar-2010, Scherer-etal-2020, Lazar-Fichtner-2021}. Kappa distributions are the result of such combined effects, both of particle-particle collisions conditioning a
quasi-Maxwellian profile at low energies, and of particle interactions with wave turbulence and fluctuations, which can explain the suprathermal tails of these distributions
\citep{Vocks-Mann-2003,Yoon-2014}.
The transport theory for such plasma populations must therefore rely on a kinetic approach centered on Boltzmann transport equation (BTE) that describes the time/space
evolution of particle velocity distribution. 
Alternatively, plasma transport theory provides a simpler, macroscopic approach to account for the moments of the velocity distributions and their variations, namely, by relating 
fluxes (e.g., heat flux or electric currents) to their sources (e.g., electromagnetic fields, gradients of temperature or density), through
linear relationships. Coefficients of proportionality are termed transport coefficients and may determine the transport of mass, momentum, and energy. \citep{Braginskii-1965,Balescu-1988,Dum-1990}.
The mathematical formalism for the transport approach used in the present work (including the simplified ansatz to allow for analytical or numerical computation of the  collision integral) is presented in  Sec.~\ref{sec:tc}.

Recently, the electric conductivity, the thermoelectric coefficient, the thermal conductivity, and the diffusion and mobility coefficients have been derived and estimated  for electron populations described by the Standard Kappa
Distribution (SKDs) \citep{Husidic-etal-2021}. Introduced in the pioneering works of \cite{Olbert-1968} and \cite{Vasyliunas-1968}, these original SKD models have the merit of allowing a direct and straightforward comparison with the quasi-thermal population at the low-energy
core of Kappa distribution, reproduced in this case by the Maxwellian limit ($\kappa \to \infty$)  \citep{Lazar-etal-2015,
Lazar-etal-2016}. Such a comparison can thus emphasise the contribution of suprathermal populations to any property of the plasma
system. \cite{Husidic-etal-2021} have shown that all the aforementioned transport coefficients are systematically and markedly
enhanced in the presence of suprathermal electrons.
Other similar studies attempting to evaluate these coefficients for SKD electrons \citep{Wang-Du-2017, Abbasi-etal-2017,
Guo-Du-2019} have in general adopted variants of Kappa distributions, which provide only underestimates of transport coefficients
\citep{Husidic-etal-2021}.

However, even in the original forms, the SKDs themselves have a number of well-known limitations. Namely, SKDs do allow for finite (convergent) macroscopic velocity moments $M_l$ of order $l$  only if power-law exponents are sufficiently high, that is, $\kappa > (l+1)/2$. These limitations have been resolved by defining the regularised Kappa distributions (RKDs) \citep{Scherer-etal-2017}, see also Appendix \ref{app:dist}. Moreover, the RKDs exhibit exponential cutoffs of the suprathermal tails, able to minimise the unphysical implication of superluminal particles with speeds exceeding the speed of light in vacuum \citep{Scherer-etal-2019}.

In the present paper we compute the main transport coefficients for the electrons for the first time described by the RKDs (Sec.~\ref{sec:tc}), 
namely the diffusion coefficient (Sec.~\ref{subsec:diffusion}), the mobility coefficient (section~\ref{subsec:mobility}), the 
electric conductivity (section~\ref{subsec:sigma}), the thermoelectric coefficient 
(section~\ref{subsec:alpha}), and the thermal conductivity (section~\ref{subsec:lambda}). 
The transport coefficients are well defined, taking finite values for any value of the power exponent $\kappa > 0$, and are not affected by any limitation given by the singularities of SKDs, e.g., for low values of $\kappa \leqslant (l+1)/2$ \citep{Lazar-etal-2020}. Conclusions and an outlook for potential future work are formulated in Sec.~\ref{sec:conclusions}. 
In the appendix we briefly discuss the RKD (Appendix~\ref{app:dist}), and give useful formulas and solutions of the integrals (Appendix~\ref{app:formulas}) occurring in Sec.~\ref{sec:tc}. 
Furthermore, in Appendix~\ref{app:results} we present approximations that allow to extend the scope of transport coefficients even to $\kappa \to 0$, and tabulated in Table~\ref{tab:results} the expressions obtained for the transport coefficients for a suggestive comparison on different distribution functions.

\section{Transport coefficients}\label{sec:tc}

Within transport theory, we may start from the velocity moments of the BTE and use macroscopic laws for the electric field, the electric current density, the heat flux and the particle flux. Comparisons between the moment equations and macroscopic laws allow us to identify the transport coefficients and to derive their expressions.
In order to study the effects of suprathermal particles on the transport coefficients, we assume heavy and stationary ions and mobile electrons described by the RKD.

The macroscopic relationships between fluxes and their sources used in the present work are
\begin{align}
    \vec{\Gamma} &= -D\,\nabla n - \mu\,n\,\vec{E}\,, \label{eq:fick's_law} \\
    \vec{\Gamma} &= \langle \vec{v} \rangle = \int \mathrm{d}^3v\,\vec{v}\,f \,, \label{eq:drift_velocity}\\
    \vec{E} &= \frac{1}{\sigma}\,\vec{j} + 
    \alpha\,\nabla T\,, \label{eq:ohm's_law} \\
    \vec{j} &= q \int \mathrm{d}^3v\,\vec{v}\,f\,, \label{eq:current_density} \\
    \vec{q} &= (\phi + \alpha\,T)\,\vec{j} 
    - \lambda\, \nabla T\,, \label{eq:fourier's_law} \\
    \vec{q} &= \frac{1}{2}m \int \mathrm{d}^3v\,v^2\,\vec{v}\,f
    = \int \mathrm{d}^3v\,\varepsilon\,\vec{v}\,f\,. \label{eq:heat_flux} 
\end{align}
The particle flux density $\vec{\Gamma}$ defined as the average of velocity $\vec{v}$ (Eq.~\ref{eq:drift_velocity}), can occur due to a gradient in number density $n$ or the presence of an electric field $\vec{E}$ in an extended Fick’s law in Eq.~\eqref{eq:fick's_law}. The corresponding transport coefficients are the diffusion coefficient $D$ and the mobility coefficient $\mu$, respectively. 
Equation~\eqref{eq:ohm's_law} is a generalised Ohm's law and sets the electric field in relation to electric current density $\vec{j}$, defined in Eq.~\eqref{eq:current_density} with electric charge $q$, and gradient in temperature $T$. There, the related transport coefficients are electric conductivity $\sigma$ and thermoelectric coefficient $\alpha$, respectively. The heat flux $\vec{q}$ as defined in Eq.~\eqref{eq:heat_flux} can arise due to an electric current density or a temperature gradient, expressed through an extended Fourier’s law in Eq.~\eqref{eq:fourier's_law}, where $\lambda$ is the thermal conductivity, $\phi$ is the electric potential related to the electric field via $\vec{E} = - \nabla \phi$, and $\alpha$ is the same thermoelectric coefficient as in Eq.~\eqref{eq:ohm's_law}. More details can be found in \cite{Spatschek-1990}, \cite{Boyd-Sanderson-2003}, and \cite{Goedbloed-etal-2019}.

The evolution of a distribution function $f$ in time and space is given by the partial differential equation
\begin{equation}\label{eq:boltzmann_eq}
    \frac{\partial f}{\partial t} + \vec{v} \cdot 
    \nabla f + \frac{q}{m} 
    \left[\vec{E} + \frac{1}{c} (\vec{v} \times \vec{B}) \right] 
    \cdot \nabla_{\vec{v}} f = \mathcal{C}(f)\,,
\end{equation}
which is called the BTE. Here, we assume that the electric and magnetic fields
$\vec{E}$ and $\vec{B}$ contain both the imposed and self-generated fields. While $\nabla$ is the standard spatial
derivative, $\nabla_{\vec{v}}$ expresses the velocity gradient. The collision term is denoted by $\mathcal{C}(f)$. Assuming stationary transport and neglecting the magnetic field in Eq.~\eqref{eq:boltzmann_eq}, we find 
\begin{equation}\label{eq:boltzmann__eq_simplified}
    \vec{v} \cdot \nabla f - \frac{q}{m}\,
    \vec{E} \cdot \nabla_{\vec{v}} f = \mathcal{C}(f)\,.
\end{equation}
Collisions between particles cause changes in the velocity distribution. Assuming that the changes are relatively small, we can linearise the distribution function and write it as a sum of the stationary solution $f_0$  and a small perturbation $f_1$, which yields
\begin{equation}\label{eq:f_linearized}
    f(\vec{r},\vec{v},t) = f_0(\vec{r},\vec{v}) + f_1(\vec{r},\vec{v},t)\,.
\end{equation}
The collision term in the BTE is in its most general form an integral that proves to be very challenging to exactly compute. To overcome this issue, we use a Krook-type collisional operator \citep{Bhatnagar-etal-1954} for $\mathcal{C}(f)$, given by
\begin{equation}\label{eq:krook_collision_term}
    \mathcal{C}(f) = - \nu_\mathrm{ei}(v)\,(f - f_0) = - \nu_\mathrm{ei}(v)\,f_1\,.
\end{equation}
This ansatz assumes that the perturbed distribution function $f$ relaxes toward the stationary solution $f_0$ under the effect of collisions that occur with frequency $\nu_\mathrm{ei}(v)$ as a function of speed $v$. Here, the subscript 'ei' indicates that collisions occur only between electrons and stationary ions. For $\nu_\mathrm{ei}(v)$ we used the expression given by \cite{Helander-Sigmar-2005}

\begin{equation}\label{eq:collision_frequency}
    \nu_\mathrm{ei}(v) = \nu_\mathrm{ei} = 
    \frac{4\,\pi\,n_\mathrm{e}\,z\,e^4\,L^\mathrm{ei}}{m_\mathrm{r}\,m_\mathrm{e}\,v^3} 
    \approx
    \frac{4\,\pi\,n_\mathrm{e}\,z\,e^4\,L^\mathrm{ei}}{m_\mathrm{e}^2\,v^3}
\end{equation}
with electron number density $n_\mathrm{e}$, ion charge number $z$, elementary charge $e$, electron mass $m_\mathrm{e}$, and reduced mass $m_\mathrm{r} \equiv 
m_\mathrm{e}\,m_\mathrm{i}/(m_\mathrm{e}+m_\mathrm{i}) \simeq m_\mathrm{e}$ (with ion mass $m_\mathrm{i}$), and Coulomb logarithm $L^\mathrm{ei} = \ln{\Lambda}$, where $\Lambda$ is the (normalised) electron Debye length. In the following, we omit the subscript e in $n$ and $m$ as they always refer to electrons for the remaining part. We further note that considering only collisions between electrons and ions is a generic choice, relevant enough for low heliocentric distances, e.g., in the outer corona where $T_\mathrm{e} < T_\mathrm{i}$ is observed \citep{Landi-2007, Landi-Cranmer-2009}, while for larger distances, where $T_\mathrm{e} \sim T_\mathrm{i}$, electron-electron collisions with frequency $\nu_{\rm ee} \gtrsim \nu_{\rm ei}$ must also be taken into account.

By inserting the linearised distribution function from Eq.~\eqref{eq:f_linearized} into the simplified BTE from Eq.~\eqref{eq:boltzmann__eq_simplified}, we obtain
\begin{equation}\label{eq:boltzmann_with_coll_freq}
     \vec{v} \cdot \nabla f_\mathrm{RKD} - \frac{e}{m}\,
    \vec{E} \cdot \nabla_{\vec{v}} f_\mathrm{RKD} = -\nu_\mathrm{ei}\,f_1\,,
\end{equation}
where we set $f_0$ to the RKD $f_\mathrm{RKD}$ and neglected all second-order terms of the spatial and velocity gradients of perturbation $f_1$. 
For the derviation of the transport coefficients we did not use the RKD in its original representation displayed in Eq.~\eqref{eq:rkd_general}, but rewrote it in terms of kinetic energy $\varepsilon = m\,v^2 /2$ and corresponding Maxwellian temperature $T$ to
\begin{equation}\label{eq:rkd_energy}
    f_{\mathrm{RKD}} =  N_\mathrm{RKD}\,
    \left(1 + \frac{\varepsilon}{\kappa\,k_\mathrm{B}\,T} \right)^{-(\kappa + 1)} 
    \exp\left(-\xi^2\frac{\varepsilon}{k_\mathrm{B}\,T}\right)
\end{equation}
with Boltzmann's constant $k_\mathrm{B}$, a dimensionless cutoff-parameter $\xi$ (see Appendix~\ref{app:dist}), and normalisation constant
\begin{equation}\label{eq:normalization_constant}
    N_\mathrm{RKD} = \frac{n}{{\pi^{3/2}\,\kappa^{3/2}}\,\mathcal{U}_0} \left(\frac{m}{2\,k_\mathrm{B}\,T}\right)^{3/2}\,,
\end{equation}
where $\mathcal{U}_0 \equiv \mathcal{U}(3/2,3/2 - \kappa,\xi^2 \kappa)$, $\mathcal{U}(a,b,x)$ being Kummer's function (see Appendices~\ref{app:dist} and \ref{app:formulas}).
Furthermore, $\kappa$ is a free parameter characterising the suprathermal tails of the distribution function.
Using this alternative form of $f_\mathrm{RKD}$, we can rewrite Eq.~\eqref{eq:boltzmann_with_coll_freq} in terms of $f_1$ to find
\begin{align}\label{eq:f_1}
    f_1 &= 
    - \frac{f_\mathrm{RKD}\,(\kappa + 1)}{\nu_\mathrm{ei}\,\left(k_\mathrm{B}\,T\,\kappa + \varepsilon\right)}
    \left(e\,\vec{E} + \frac{\varepsilon}{T}\, \nabla T\right) \cdot \vec{v} \nonumber \\
    &- \frac{f_\mathrm{RKD}\,\xi^2}{\nu_\mathrm{ei}\,k_\mathrm{B}\,T}
    \left(e\,\vec{E} + \frac{\varepsilon}{T}\, \nabla T\right) \cdot \vec{v} \nonumber \\
    &+ \frac{3}{2} \frac{f_\mathrm{RKD}}{\nu_\mathrm{ei}} \frac{\nabla T}{T} \cdot \vec{v} 
    - \frac{f_\mathrm{RKD}}{\nu_\mathrm{ei}\,n} \nabla n \cdot \vec{v}\,.
\end{align}
%

\subsection{Diffusion coefficient} \label{subsec:diffusion}
The diffusion coefficient $D$ is derived by setting $\vec{E} = \vec{0}$ and $\nabla T = \vec{0}$.
Equation~\eqref{eq:f_1} then reduces to  
\begin{equation}\label{eq:f_1_for_diffusion}
    f_1 = - \frac{f_\mathrm{RKD}}{\nu_\mathrm{ei}\,n} \nabla n \cdot \vec{v}   \,.
\end{equation}
The particle flux in Eq.~\eqref{eq:drift_velocity} simplifies to
\begin{equation}\label{eq:diffusion_integral_general}
    \vec{\Gamma} = \int \mathrm{d}^3v\,\vec{v}\,f 
    = \int \mathrm{d}^3v\,\vec{v}\,f_1\,,
\end{equation}
as $\int \mathrm{d}^3v\,\vec{v}\,f_\mathrm{RKD} = 0$ due to the odd integrand with respect to the velocity. 
Inserting Eq.~\eqref{eq:f_1_for_diffusion} into \eqref{eq:diffusion_integral_general} yields
\begin{equation}\label{eq:drift_velocity_with_average_for_diffusion}
    \vec{\Gamma}= - \frac{1}{n}\, \int \mathrm{d}^3v\,
    \frac{\vec{v} \vec{v}}{\nu_\mathrm{ei}} f_\mathrm{RKD} \nabla n = - \frac{1}{3\,n}\,
    \Bigg\langle \frac{v^2}{\nu_\mathrm{ei}} \Bigg\rangle \,\nabla n\,.
\end{equation}
The vanishing cross diagonal terms in the dyadic product $\vec{v}\vec{v}$ in Eq.~\eqref{eq:drift_velocity_with_average_for_diffusion} allow to rewrite the integral into
$\int \mathrm{d}^3v\, \vec{v} \vec{v}\,G(v) 
= 1/3\,\mathcal{I} \int \mathrm{d}^3v\,v^2\,G(v)$, where $\mathcal{I}$ denotes the unit tensor and $G(v)$ is some function of speed $v$. Then the integral can be transformed into an average value $\langle F(v) \rangle \equiv \int \mathrm{d}^3v\,F(v)\,f_\mathrm{RKD}$, where $F(v)$ is some function of $v$. This procedure is performed for all considered transport coefficients in the present work.

By comparing Eqs.~\eqref{eq:drift_velocity_with_average_for_diffusion} and \eqref{eq:fick's_law} we can identify the diffusion coefficient as
\begin{equation}\label{eq:D_with_average}
    D = \frac{1}{3\,n}\,
    \Bigg\langle \frac{v^2}{\nu_\mathrm{ei}} \Bigg\rangle\
    = \frac{m^2}{12\,\pi\,z\,e^4\,L^\mathrm{ei}\,n^2}\, \Big \langle v^5 \Big \rangle\,.
\end{equation}
\cite{Scherer-etal-2020} derived general solutions of integrals that contain (regularised) Kappa distributions (see also Appendix~\ref{app:formulas}), and after solving the integral we obtain the diffusion coefficient based on the RKD as
\begin{equation}\label{eq:diffusion_coefficient}
    D= \underbrace{\frac{4\,\sqrt{2}\,(k_\mathrm{B}\,T)^{5/2}}{m^{1/2}\,\pi^{3/2}\,n\,z\,e^4\,L^\mathrm{ei}}}_\text{$\equiv D_\mathrm{M}$}\, \frac{\kappa^{5/2}\,
    \mathcal{U}_{[4,4]}}{\mathcal{U}_0}
    = D_\mathrm{M} \frac{\kappa^{5/2}\,
    \mathcal{U}_{[4,4]}}{\mathcal{U}_0}\,,
\end{equation}
where $D_{M}$ is the Maxwellian diffusion coefficient and where we made use of the compact notation given in Eq~\eqref{eq:compact_kummer}, that is, $\mathcal{U}_{[l,m]} \equiv \mathcal{U} \left(l, m - \kappa, \xi^2 \kappa \right)$.
By setting $\xi = 0$, Eq.~\eqref{eq:diffusion_coefficient} becomes the diffusion coefficient based on the SKD (see \cite{Husidic-etal-2021} for all transport coefficients based on the SKD) as
\begin{equation}\label{eq:diffusion_coefficient_skd}
    D =  D_\mathrm{M}  \frac{\Gamma(\kappa - 3)}{\Gamma(\kappa - 1/2)}\,\kappa^{5/2}\,.
\end{equation}
From Eq.~\eqref{eq:diffusion_coefficient} (as well as from all the subsequent transport coefficients below) we can see that the diffusion coefficient can be written as a product of a Maxwellian part and a $\kappa$-dependent part. This is a consequence of the composition of the RKD (see \cite{Scherer-etal-2020} for a detailed discussion), and enables a simple assessment of the influence of suprathermal particles on the transport coefficients.

\begin{figure}
\includegraphics[width=\hsize]{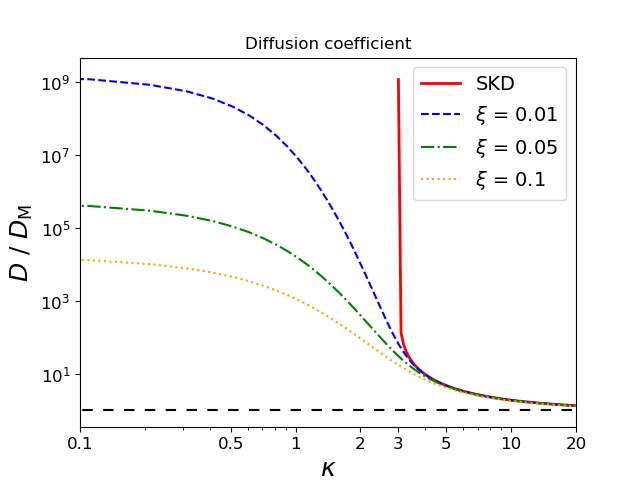}
   \caption{The plot displays the diffusion coefficient $D$ as a function of $\kappa$. The curves show the results based on the SKD and three RKDs with different cutoff parameters (see legend). All functions are normalised to the Maxwellian limit (dashed horizontal line).}
              \label{fig:diffusion}%
    \end{figure}

Figure~\ref{fig:diffusion} shows the diffusion coefficient $D$ as a function of $\kappa$ and based on the SKD and three RKDs with different cutoff parameters (see legend). While the SKD-based result diverges approaching $\kappa = 3$, the RKD-based results resolve the singularity, allowing for a continuation to $\kappa < 3$. Furthermore, for increasing values of $\xi$, the values for $D$ become smaller. The maximum value of the diffusion coefficient as well as for all other transport coefficients are presented in Appendix~\ref{app:results}.

%

\subsection{Mobility coefficient} \label{subsec:mobility}
The mobility coefficient $\mu$ in Eq.~\eqref{eq:fick's_law} is obtained by setting $\nabla n = \vec{0}$ and $\nabla T = \vec{0}$. Equation~\eqref{eq:f_1} then becomes
\begin{equation}\label{eq:f_1_for_mobility}
    f_1 = 
    \left[- \frac{f_\mathrm{RKD}\,(\kappa + 1)\,e}{\nu_\mathrm{ei}\,\left(k_\mathrm{B}\,T\,\kappa + \varepsilon\right)} - \frac{f_\mathrm{RKD}\,\xi^2\,e}{\nu_\mathrm{ei}\,k_\mathrm{B}\,T} \right]
    \vec{E} \cdot \vec{v}\,.
\end{equation}
We can continue from Eq.~\eqref{eq:diffusion_integral_general} and insert Eq.~\eqref{eq:f_1_for_mobility} to obtain
\begin{align}\label{eq:current_density_with_average_for_mobility}
    \vec{\Gamma} = &-(\kappa + 1)\,e\,\int \mathrm{d}^3v\,
    \frac{\vec{v} \vec{v}}{\nu_\mathrm{ei}} 
    \left(k_\mathrm{B}\,T\,\kappa + \varepsilon \right)^{-1} f_\mathrm{RKD} \,\vec{E}  \nonumber \\
    &- \frac{\xi^2\,e}{k_\mathrm{B}\,T}\,\int \mathrm{d}^3v\,
    \frac{\vec{v} \vec{v}}{\nu_\mathrm{ei}} 
     f_\mathrm{RKD} \,\vec{E} \nonumber \\
    &= \Bigg[ -\frac{(\kappa + 1)\,e}{3}\,
    \Bigg\langle \frac{v^2}{\nu_\mathrm{ei}}
    \left(k_\mathrm{B}\,T\,\kappa + \varepsilon \right)^{-1} \Bigg\rangle  \nonumber \\
    &- \frac{\xi^2\,e}{3\,k_\mathrm{B}\,T}
    \Bigg\langle \frac{v^2}{\nu_\mathrm{ei}} \Bigg\rangle \Bigg] \,\vec{E}\,.
\end{align}
Comparing Eqs.~\eqref{eq:current_density_with_average_for_mobility} and \eqref{eq:fick's_law} allows to identify the mobility coefficient as
\begin{equation}\label{eq:mu_with_average}
  \begin{split}
    \mu &= \frac{(\kappa + 1)\,e}{3\,n}\,
    \Bigg\langle \frac{v^2}{\nu_\mathrm{ei}} 
    \left(k_\mathrm{B}\,T\,\kappa + \varepsilon \right)^{-1} \Bigg\rangle 
    + \frac{\xi^2\,e}{3\,k_\mathrm{B}\,T\,n}
    \Bigg\langle \frac{v^2}{\nu_\mathrm{ei}} \Bigg\rangle \,.
  \end{split}
\end{equation}
We insert the expression for the collision frequency from Eq.~\eqref{eq:collision_frequency} and introduce the quantity $A_\kappa \equiv m/(k_\mathrm{B}\,T\,\kappa)$, which leads to
\begin{align}\label{eq:mu_before_int}
        \mu = & \frac{(\kappa + 1)\,m^2}{12\,\pi\,n^2\,k_\mathrm{B}\,T\,\kappa\,z\,e^3\,L_\mathrm{ei}}\,
        \Bigg\langle v^5\,\left(1 + A_\kappa\,v^2 \right)^{-1}  \Bigg\rangle \nonumber \\
        & + \frac{\xi^2\,m^2}{12\,\pi\,n^2\,k_\mathrm{B}\,T\,z\,e^3\,L_\mathrm{ei}}\,
        \Big\langle v^5  \Big\rangle\,.
\end{align}
After solving the integrals (see Appendix~\ref{app:formulas}), the mobility coefficient based on the RKD takes the form
\begin{align}\label{eq:mobility_coefficient}
        \mu = & \underbrace{\frac{4\,\sqrt{2}\,(k_\mathrm{B}\,T)^{3/2}}{m^{1/2}\,\pi^{3/2}\,n\,z\,e^3\,L}}_\text{$\equiv \mu_\mathrm{M}$}\, \frac{\kappa^{3/2}}{\mathcal{U}_0}
    \left[(\kappa+1)\,\mathcal{U}_{[4,3]} + \kappa\,\xi^2\,\mathcal{U}_{[4,4]} \right]  \nonumber \\ 
    = & \mu_\mathrm{M} \, \frac{\kappa^{3/2}}{\mathcal{U}_0}
    \left[(\kappa+1)\,\mathcal{U}_{[4,3]} + \kappa\,\xi^2\,\mathcal{U}_{[4,4]} \right]\,,
\end{align}
where $\mu_\mathrm{M}$ is the Maxwellian mobility coefficient.
With $\xi = 0$, Eq.~\eqref{eq:mobility_coefficient} becomes the mobility coefficient based on the SKD
\begin{equation}\label{eq:mobility_coefficient_skd}
    \mu = \mu_\mathrm{M} \frac{\Gamma\left(\kappa - 2\right)}{\Gamma\left(\kappa - 1/2\right)}\, \kappa^{3/2}\,.
\end{equation}
The Einstein equation, which establishes the relationship between diffusion and mobility coefficients, is obtained by combining Eqs.~\eqref{eq:diffusion_coefficient} and \eqref{eq:mobility_coefficient}, yielding
\begin{equation}\label{eq:einstein_relation}
    D = \frac{\kappa\,\mathcal{U}_{[4,4]}}{(\kappa+1)\,\mathcal{U}_{[4,3]} + \kappa\,\xi^2\,\mathcal{U}_{[4,4]}}\,\frac{ \mu\,k_\mathrm{B}\,T}{e}\,.
\end{equation}
By setting $\xi = 0$, this equation can be simplified to the case for the SKD
\begin{equation}\label{eq:einstein_relation_skd}
    D = \frac{\kappa}{\kappa - 3}\,\frac{ \mu\,k_\mathrm{B}\,T}{e}\,,
\end{equation}
and with $\kappa \to \infty$ further simplified to the well-known Maxwellian-based result $D_\mathrm{M} = \mu_\mathrm{M}\,k_\mathrm{B}\,T/e$.
   \begin{figure}
   \includegraphics[width=\hsize]{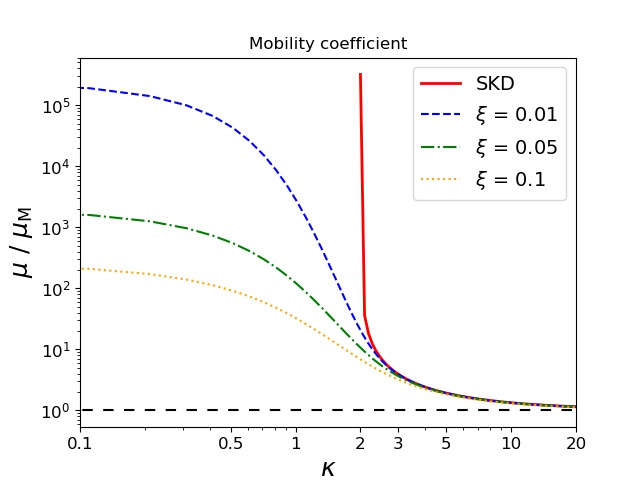}
   \caption{The plot displays the mobility coefficient $\mu$ as a function of $\kappa$. The curves show the results based on the SKD and three RKDs with different cutoff parameters (see legend). All functions are normalised to the Maxwellian limit (dashed horizontal line).}
              \label{fig:mobility}%
    \end{figure}
The estimates of the mobility coefficient for three RKDs with different cutoff parameters, and the corresponding SKD are displayed in Fig.~\ref{fig:mobility}. The result based on the SKD diverges at $\kappa = 2$, whereas the RKD-based results continue to lower values of $\kappa$. By increasing the cutoff parameter $\xi$, the value of the mobility coefficient becomes smaller.

\subsection{The electric conductivity} \label{subsec:sigma}
Similarly to the mobility coefficient, we set $\nabla n = \vec{0}$ and $\nabla T = \vec{0}$ to calculate the electric conductivity. We find for $f_1$ the same expression 
as in Eq.~\eqref{eq:f_1_for_mobility}, which we then insert in Eq.~\eqref{eq:current_density} to find
\begin{equation}\label{eq:current_density_with_f_1_for_sigma}
    \vec{j} = -e \int \mathrm{d}^3v\,\vec{v}\,f
    = -e \int \mathrm{d}^3v\,\vec{v}\,f_1\,.
\end{equation}
Equation~\eqref{eq:current_density_with_f_1_for_sigma} differs from 
Eq.~\eqref{eq:diffusion_integral_general} only in an additional factor $-e$, leading to the same integrals as in Eq.~\eqref{eq:current_density_with_average_for_mobility}. Thus, using Eq.~\eqref{eq:ohm's_law} we can immediately write the electric conductivity based on the RKD as
\begin{align}\label{eq:sigma}
        \sigma = & \underbrace{\frac{4\,\sqrt{2}\,(k_\mathrm{B}\,T)^{3/2}}{m^{1/2}\,\pi^{3/2}\,z\,e^2\,L}}_\text{$\equiv \sigma_\mathrm{M}$}\, \frac{\kappa^{3/2}}{\mathcal{U}_0}
    \left[(\kappa+1)\,\mathcal{U}_{[4,3]} + \kappa\,\xi^2\,\mathcal{U}_{[4,4]} \right]
    \nonumber \\
    = &\sigma_\mathrm{M} \, \frac{\kappa^{3/2}}{\mathcal{U}_0}
    \left[(\kappa+1)\,\mathcal{U}_{[4,3]} + \kappa\,\xi^2\,\mathcal{U}_{[4,4]} \right]
\end{align}
with $\sigma_\mathrm{M}$ being the Maxwellian electric conductivity.
The relation between $\sigma$ and $\mu$ is
\begin{equation}\label{eq:relation_mu_sigma}
    \sigma = n\,e\,\mu\,.
\end{equation}
We set $\xi = 0$ to obtain the SKD-based result for $\sigma$ with
\begin{equation}\label{eq:sigma_skd}
   \sigma = \sigma_\mathrm{M}
   \frac{\Gamma\left(\kappa - 2\right)}{\Gamma\left(\kappa - 1/2\right)}\, \kappa^{3/2}\,.
\end{equation}
   \begin{figure}
   \includegraphics[width=\hsize]{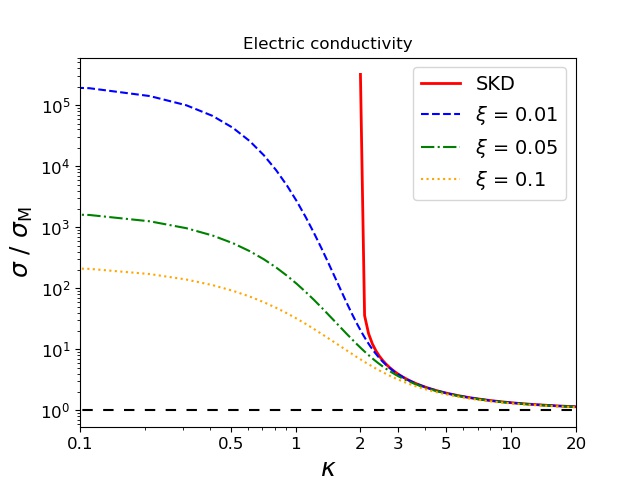}
   \caption{The plot displays the electric conductivity $\sigma$ as a function of $\kappa$. The curves show the results based on the SKD and three RKDs with different cutoff parameters (see legend). All functions are normalised to the Maxwellian limit (dashed horizontal line).}
              \label{fig:sigma}%
    \end{figure}

Figure~\ref{fig:sigma} shows the electric conductivity as a function of $\kappa$ for three different RKDs and, to compare, the corresponding SKD. Since $\sigma/\sigma_{\rm M} = \mu / \mu_{\rm M}$ (see Eq.~\ref{eq:mobility_coefficient}), the relative values obtained in this case are exactly the same with those displayed for the mobility coefficient in Figure~\ref{fig:mobility}.

\subsection{The thermoelectric coefficient} \label{subsec:alpha}
The thermoelectric coefficient is derived by setting $\vec{E} = \vec{0}$ and $\nabla n = \vec{0}$, which 
simplifies Eq.~\eqref{eq:f_1} to
\begin{equation}\label{eq:f_1_for_alpha}
    f_1 = 
    \left[- \frac{f_\mathrm{RKD}\,(\kappa + 1)\,\varepsilon}{\nu_\mathrm{ei}\,\left(k_\mathrm{B}\,T\,\kappa + \varepsilon\right)}
    - \frac{f_\mathrm{RKD}\,\xi^2\,\varepsilon}{k_\mathrm{B}\,T\,\nu_\mathrm{ei}}
    + \frac{3}{2} \frac{f_\kappa}{\nu_\mathrm{ei}}\right] \frac{\nabla T}{T} \cdot \vec{v}\,.
\end{equation}
We insert Eq.~\eqref{eq:f_1_for_alpha} into \eqref{eq:current_density_with_f_1_for_sigma} to obtain
\begin{align}\label{eq:current_density_for_alpha}
    \vec{j} &=  \frac{(\kappa + 1)\,e}{T} \nabla T \cdot 
    \int \mathrm{d}^3v\,
    \frac{\vec{v}\vec{v}}{\nu_\mathrm{ei}} 
    \left(k_\mathrm{B}\,T\,\kappa + \varepsilon \right)^{-1} \varepsilon\, f_\kappa 
    \nonumber \\ & + \frac{\xi^2\,e}{k_\mathrm{B}\,T^2} \nabla T \cdot \int \mathrm{d}^3v\,
    \frac{\vec{v} \vec{v}}{\nu_\mathrm{ei}}\,\varepsilon\, f_\kappa \nonumber \\
    &- \frac{3}{2} \frac{e}{T} \nabla T \cdot \int \mathrm{d}^3v\,
    \frac{\vec{v} \vec{v}}{\nu_\mathrm{ei}} f_\kappa \nonumber \\
    &= \Bigg[\frac{(\kappa + 1)\,e}{3\,T} 
    \Bigg\langle \frac{v^2\,\varepsilon}{\nu_\mathrm{ei}} 
    \left(k_\mathrm{B}\,T\,\kappa + \varepsilon \right)^{-1} \Bigg\rangle \nonumber \\
    &+ \frac{\xi^2\,e}{3\,k_\mathrm{B}\,T^2} \Bigg\langle \frac{v^2\,\varepsilon}{\nu_\mathrm{ei}} \Bigg\rangle 
    - \frac{e}{2\,T} \Bigg\langle \frac{v^2}{\nu_\mathrm{ei}} \Bigg\rangle \Bigg] \nabla T\,.
\end{align}
By comparing the coefficients in Eq.~\eqref{eq:current_density_for_alpha} and \eqref{eq:ohm's_law} we are able to identify the thermoelectric coefficient as
\begin{align}\label{eq:alpha_with_average}
    \alpha &= -\frac{(\kappa + 1)\,e}{3\,T\,\sigma} 
    \Bigg\langle \frac{v^2\,\varepsilon}{\nu_\mathrm{ei}} 
    \left(k_\mathrm{B}\,T\,\kappa + \varepsilon \right)^{-1} \Bigg\rangle \nonumber \\
    &- \frac{\xi^2\,e}{3\,k_\mathrm{B}\,T^2\,\sigma} \Bigg\langle \frac{v^2\,\varepsilon}{\nu_\mathrm{ei}} \Bigg\rangle 
    + \frac{e}{2\,T\,\sigma} \Bigg\langle \frac{v^2}{\nu_\mathrm{ei}} \Bigg\rangle \,.
\end{align}
We insert the expression for the collision frequency from Eq.~\eqref{eq:collision_frequency} into the equation above to obtain
\begin{align}\label{eq:alpha_before_int}
        \alpha = & - \frac{(\kappa + 1)\,m^3}{24\,\pi\,n\,z\,e^3\,k_\mathrm{B}\,T^2\,\kappa\,L_\mathrm{ei}\,\sigma}
        \Bigg\langle v^7\,\left(1 + A_\kappa\,v^2 \right)^{-1}  \Bigg\rangle         \nonumber \\
        & - \frac{\xi^2\,m^3}{24\,\pi\,n\,z\,e^3\,k_\mathrm{B}\,T^2\,L_\mathrm{ei}\,\sigma}
        \Big\langle v^7  \Big\rangle \nonumber \\
        & + \frac{m^2}{8\,\pi\,n\,z\,e^3\,T\,L_\mathrm{ei}\,\sigma} \Big\langle v^5  \Big\rangle\,.
\end{align}
Solving the integrals and inserting the found expression for $\sigma$ from Eq.~\eqref{eq:sigma} yields the thermoelectric coefficient based on the RKD in the form
\begin{align}\label{eq:alpha}
    \alpha = & \underbrace{-\frac{5\,k_\mathrm{B}}{2\,e}}_\text{$\equiv \alpha_\mathrm{M}$}
    \frac{\kappa\,\left[4\,(\kappa+1)\,\mathcal{U}_{[5,4]} -\frac{3}{2}\,
    \mathcal{U}_{[4,4]}
    + 4\,\kappa\,\xi^2\,
    \mathcal{U}_{[5,5]}\right]}{5/2\,\left[(\kappa+1)\,\mathcal{U}_{[4,3]} + \kappa\,\xi^2\,\mathcal{U}_{[4,4]}\right]}\nonumber \\
    = & \alpha_\mathrm{M}\,
    \frac{\kappa\,\left[4\,(\kappa+1)\,\mathcal{U}_{[5,4]} -\frac{3}{2}\,
    \mathcal{U}_{[4,4]}
    + 4\,\kappa\,\xi^2\,
    \mathcal{U}_{[5,5]}\right]}{5/2\,\left[(\kappa+1)\,\mathcal{U}_{[4,3]} + \kappa\,\xi^2\,\mathcal{U}_{[4,4]}\right]}
\end{align}
with the Maxwellian thermoelectric coefficient $\alpha_\mathrm{M}$.
Furthermore, with $\xi = 0$ the SKD-based result for $\alpha$ becomes
\begin{equation}\label{eq:alpha_skd}
        \alpha = \alpha_\mathrm{M} \frac{\kappa}{\kappa - 3}\,.
\end{equation}

\begin{figure}
   \includegraphics[width=\hsize]{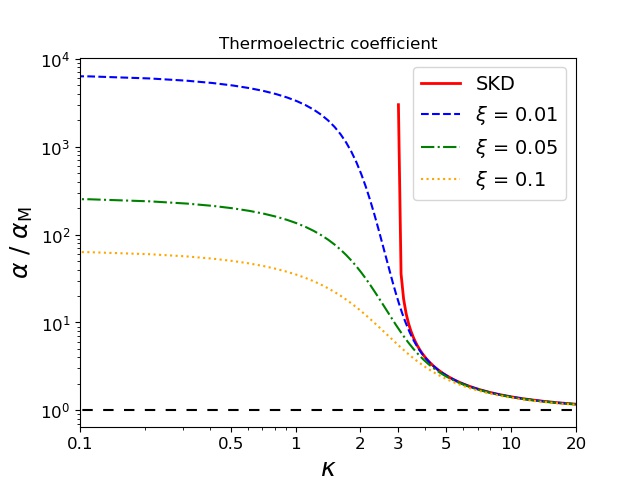}
   \caption{The plot displays the thermoelectric coefficient $\alpha$ as a function of $\kappa$. The results are based on the SKD and three RKDs with different cutoff parameters (see legend). All functions are normalised to the Maxwellian limit (dashed horizontal line).}
              \label{fig:alpha}%
    \end{figure}

Figure~\ref{fig:alpha} displays the thermoelectric coefficient as a function of $\kappa$. Similarly to the previous transport coefficients, we see that the SKD-based result has a singularity (here at $\kappa = 3$), which is resolved by the RKD.

\vspace{0.1cm}

\subsection{The thermal conductivity} \label{subsec:lambda}
For the the derivation of the thermal conductivity we set $\nabla n = \vec{0}$ and assume the absence of an electric current ($\vec{j} = \vec{0}$), which simplifies Eq.~\eqref{eq:ohm's_law} to $\vec{E} = \alpha\,\nabla T$.
Equation~\eqref{eq:f_1} then becomes
\begin{align}\label{eq:f_1_for_lambda}
    f_1 &=
    - \frac{f_\kappa\,(\kappa + 1)\,(\alpha\,e + \varepsilon/T)}
    {\nu_\mathrm{ei}\,\left(k_\mathrm{B}\,T\,\kappa + \varepsilon\right)}
    \, \nabla T \cdot \vec{v} \nonumber \\
    &- \frac{f_\kappa\,\xi^2\,(\alpha\,e + \varepsilon/T)}{\nu_\mathrm{ei}\,k_\mathrm{B}\,T} \nabla T \cdot \vec{v}
    + \frac{3}{2} \frac{f_\kappa}{\nu_\mathrm{ei}} \frac{\nabla T}{T} \cdot \vec{v}\,,
\end{align}
which we insert into Eq.~\eqref{eq:heat_flux} to obtain
\begin{align}\label{eq:heat_flux_for_lambda}
        \vec{q} &= -(\kappa + 1)\, \int \mathrm{d}^3v\,
    \frac{\vec{v} \vec{v}}{\nu_\mathrm{ei}} 
    \frac{(\alpha\,e + \varepsilon/T)\,\varepsilon\,f_\kappa}{k_\mathrm{B}\,T\,\kappa + \varepsilon}\,\nabla T \nonumber \\
    &- \frac{\xi^2}{k_\mathrm{B}\,T} \int \mathrm{d}^3v\,
    \frac{\vec{v} \vec{v}}{\nu_\mathrm{ei}} (\alpha\,e + \varepsilon/T)\, \varepsilon\,f_\kappa\,\nabla T \nonumber \\
    &+ \frac{3}{2\,T} \int \mathrm{d}^3v\,
    \frac{\vec{v} \vec{v}}{\nu_\mathrm{ei}} \varepsilon\,f_\kappa\,\nabla T \nonumber \\
    &= \Bigg[- \frac{\kappa + 1}{3} \Bigg\langle \frac{v^2}{\nu_\mathrm{ei}} 
    \frac{(\alpha\,e + \varepsilon/T)\,\varepsilon}{k_\mathrm{B}\,T\,\kappa + \varepsilon} \Bigg\rangle \nonumber \\
    &- \frac{\xi^2}{3\,k_\mathrm{B}\,T} \Bigg\langle \frac{v^2}{\nu_\mathrm{ei}} 
    \,(\alpha\,e + \frac{\varepsilon}{T})\,\varepsilon \Bigg\rangle + \frac{1}{2\,T} \Bigg\langle \frac{v^2}{\nu_\mathrm{ei}} \varepsilon \Bigg\rangle \Bigg] \nabla T\,.
\end{align}
A comparison of the coefficients in Eqs.~\eqref{eq:heat_flux_for_lambda} and \eqref{eq:fourier's_law} leads for the thermal conductivity to the expression
\begin{align}\label{eq:lambda_with_average}
    \lambda &= \frac{\kappa + 1}{3} \Bigg\langle \frac{v^2}{\nu_\mathrm{ei}} 
    \frac{(\alpha\,e + \varepsilon/T)\,\varepsilon}{k_\mathrm{B}\,T\,\kappa + \varepsilon} \Bigg\rangle \nonumber \\
    &+ \frac{\xi^2}{3\,k_\mathrm{B}\,T} \Bigg\langle \frac{v^2}{\nu_\mathrm{ei}} 
    \,\left(\alpha\,e + \frac{\varepsilon}{T}\right)\,\varepsilon \Bigg\rangle
    - \frac{1}{2\,T} \Bigg\langle \frac{v^2}{\nu_\mathrm{ei}} \varepsilon \Bigg\rangle\,.
\end{align}
We plug the collision frequency from Eq.~\eqref{eq:collision_frequency} into the equation above to find
\begin{align}
        \lambda = & \frac{(\kappa + 1)\,\alpha\,m^3}{24\,\pi\,n\,z\,e^3\,L_\mathrm{ei}\,k_\mathrm{B}\,T\,\kappa}
        \Bigg\langle v^7\,\left(1 + A_\kappa\,v^2 \right)^{-1}  \Bigg\rangle \nonumber \\
        &+ \frac{(\kappa + 1)\,m^4}{48\,\pi\,n\,z\,e^4\,L_\mathrm{ei}\,k_\mathrm{B}\,T^2\,\kappa}
        \Bigg\langle v^9\,\left(1 + A_\kappa\,v^2 \right)^{-1}  \Bigg\rangle \nonumber \\
        &+ \frac{\xi^2\,\alpha\,m^3}{24\,\pi\,n\,z\,e^3\,L_\mathrm{ei}\,k_\mathrm{B}\,T}
        \Big\langle v^7  \Big\rangle 
        + \frac{\xi^2\,m^4}{48\,\pi\,n\,z\,e^4\,L_\mathrm{ei}\,k_\mathrm{B}\,T^2}
        \Big\langle v^9  \Big\rangle \nonumber \\
        & - \frac{m^3}{16\,\pi\,n\,z\,e^4\,L_\mathrm{ei}\,T}
        \Big\langle v^7 \Big\rangle \,.
\end{align}
After solving the integrals and inserting the thermoelectric coefficient from Eq.~\eqref{eq:alpha}, we obtain the
thermal conductivity $\lambda$ based on the RKD, reading
\begin{align}\label{eq:lambda}
    \lambda &= \underbrace{\frac{16\,\sqrt{2}\,k_\mathrm{B}\,(k_\mathrm{B}\,T)^{5/2}}{m^{1/2}\,\pi^{3/2}\,z\,e^4\,L}}_\text{$\equiv \lambda_\mathrm{M}$}
    \frac{\kappa^{7/2}}{\mathcal{U}_0} 
    \Biggl[5(\kappa + 1)\,\mathcal{U}_{[6,5]}
    + 5\,\xi^2\,\kappa\,\mathcal{U}_{[6,6]} \nonumber \\
    &-\frac{3}{2}\,\mathcal{U}_{[5,5]} 
    - \left[(\kappa+1)\, \mathcal{U}_{[5,4]}
    + \xi^2\,\kappa\,\mathcal{U}_{[5,5]}\right] \nonumber \\
     &\times \frac{4\,(\kappa+1)\,\mathcal{U}_{[5,4]} -\frac{3}{2}\,
    \mathcal{U}_{[4,4]}
    + 4\,\kappa\,\xi^2\,
    \mathcal{U}_{[5,5]}}{(\kappa+1)\,\mathcal{U}_{[4,3]} + \kappa\,\xi^2\,\mathcal{U}_{[4,4]}}
     \Biggr]\,.
\end{align}
Setting $\xi = 0$ yields $\lambda$ based on the SKD as
\begin{equation}\label{eq:lambda_skd}
    \lambda = \lambda_\mathrm{M} \frac{\Gamma(\kappa - 4)}{\Gamma\left(\kappa - 1/2\right)}\,
    \frac{\kappa^{7/2}\,(\kappa - 1/2)}{(\kappa - 3)}\,.
\end{equation}
\begin{figure}[h!]
\includegraphics[width=\hsize]{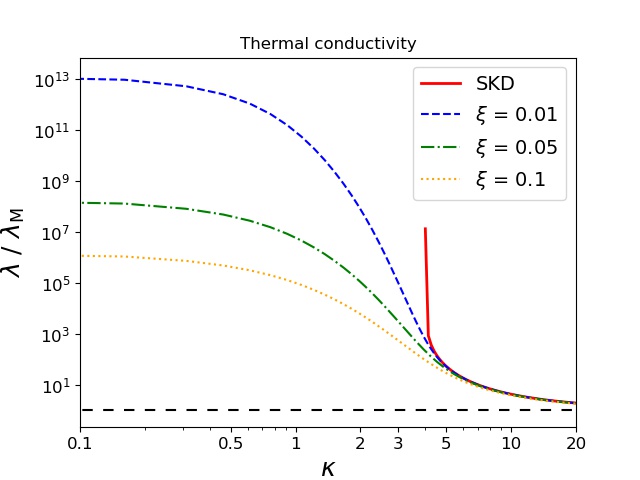}
\caption{The plot displays the thermal conductivity $\lambda$ as a function of $\kappa$. The curves show the results based on the SKD and three RKDs with different cutoff parameters (see legend). All functions are normalised to the Maxwellian limit (dashed horizontal line).}\label{fig:lambda}%
\end{figure}

Figure~\ref{fig:lambda} shows the thermal conductivity as a function of $\kappa$ and with the same composition as the previous figures. While the SKD-based result has a singularity at $\kappa = 4$, the RKD removes the pole and allows $\lambda$ to continue to values $\kappa < 4$.

\section{Conclusions and outlook} \label{sec:conclusions}
The results presented in this paper respond to the current high interest to evaluate transport coefficients in nonequilibrium space plasmas, where the effects of Coulomb collisions are counterbalanced by the interactions of plasma particles with the wave turbulence and fluctuations. It seems that this interplay can also offer a plausible explanation for the observed Kappa-like distribution of particle velocities \citep{Yoon-2011,Bian-etal-2014}, a distribution that is nearly Maxwellian at low energies, but decreases as a power-law with increasing energies (up to a few keV) \citep{Pierrard-Lazar-2010}. A macroscopic description of these plasmas depends on the nature of particle velocity distributions, and in this case should rely on Kappa distribution models. However, macroscopic velocity moments of standard Kappa distributions (SKDs) \citep{Olbert-1968, Vasyliunas-1968}, such as pressure, temperature and heat flux, cannot be defined for distributions with hard suprathermal tails, i.e., low power exponents $\kappa \leqslant 3/2$ \citep{Lazar-etal-2020}. Moreover, recent derivations of transport coefficients for SKD electrons have shown that their mobility and electric conductivity
cannot be defined for $\kappa \leqslant 2$, diffusion coefficient and thermoelectric coefficient become divergent for $\kappa \leqslant 3$, and thermal conductivity for $\kappa \leqslant 4$ \citep{Husidic-etal-2021}. 

In this paper we derived new expressions of these transport coefficients assuming the electrons described by a regularised Kappa
distribution (RKD) that has the merit to resolve all these mathematical divergences and enable a well-defined macroscopic
parameterisation \citep{Scherer-etal-2017}. It also reduces the unphysical contributions of superluminal electrons from the tails of an SKD \citep{Scherer-etal-2019}. All macroscopic parameters, including the transport
coefficients mentioned above, are found to be well defined for all values of $\kappa > 0$. Moreover, for low values of the
power exponent $\kappa$, i.e., below the SKD poles, values obtained for the transport coefficients can be markedly higher, order of magnitudes higher
than the corresponding Maxwellian limits. That means that transport coefficients can be significantly underestimated if evaluated in the
absence of suprathermal electrons. For instance, even for a moderate presence of suprathermals, i.e., for $\kappa = 2.5$, and
a fair cutoff, i.e., $\xi = 0.05$, we obtain $\mu / \mu_{\rm M} \approx 5.5$ for mobility (and the
same for electric conductivity $\sigma$), $\alpha / \alpha_{\rm M} \approx 17.6$ for thermoelectric
coefficient, and much higher differences, like, $D / D_{\rm M} \approx 93.4$ for the diffusion coefficient, 
or $\lambda / \lambda_{\rm M} \approx 1.7 \times 10^4$ for thermal conductivity.

The choice of $\xi$ in the above numerical example is somewhat arbitrary and follows mainly the requirements that $\xi > \Theta/c$ and that the essential property of Kappa distributions is retained, namely, the consideration of a sufficient number of suprathermal particles.
The sensitivity of the solutions for the transport coefficients to the value of $\xi$ becomes obvious if $\xi$ is slightly varied. For instance, if we consider the diffusion coefficient from the numerical example from above with $\kappa = 2.5$ and increase $\xi$ to 0.06, $D/D_\mathrm{M}$ is reduced by a factor of about 1.26. For smaller values of $\kappa$ this factor increases, e.g., for the limit $\kappa \to 0$, by a factor of about 2.49. However, this sensitivity is not an artefact of the RKD. It rather expresses the physical fact that
the diffusion coefficient, when calculated with the standard definition used, 
depends critically on the cut-off of a distribution function, which it must have. We note that the advantage of the RKD holds nonetheless. Using an SKD, 
$\kappa$ values below 3 - which frequently occur in the solar wind - would not be 
accessible at all, as the diffusion coefficient diverges. However, using an RKD, it is reduced to 
finite values. Furthermore, the sensitivity of the diffusion coefficient with respect to
$\kappa$ is extreme for the SKD when approaching the critical value of 3, while for the 
RKD it is much reduced and getting successively smaller with decreasing $\kappa$. 
These advantages outweigh the high sensitivity with respect to $\xi$ 
even if one would not accept it as a consequence of a distribution's cut-off. In addition, it is unclear whether observations can reveal such subtle differences in the $\xi$-parameter.

We conclude by reaffirming that based on the RKD models, realistic and physically well-defined parametrizations of the observed non-equilibrium plasmas become possible now. Future studies should confront our results with the estimations of these transport coefficients from a direct numerical integration of observational data. The best RKD fit must be conditioned only by the observed velocity distribution, without any theoretical restriction for the power exponent $\kappa$.
\newline

\section*{Acknowledgements}
The authors acknowledge support from the Ruhr-University Bochum and the Katholieke Universiteit Leuven. 
EH is grateful to the Space Weather Awareness Training Network (SWATNet) funded by the European Union’s Horizon 2020 research and innovation programme under the Marie Sk{\l}odowska-Curie grant agreement No 955620. ML and HF are grateful for support by the German Research Foundation (Deutsche Forschungsgemeinschaft, DFG) through project SCHL~201/35-1. Furthermore, these results were obtained in the framework of the projects C14/19/089  (C1 project Internal Funds KU Leuven), G.0D07.19N  (FWO-Vlaanderen), SIDC Data Exploitation (ESA Prodex-12), and Belspo projects BR/165/A2/CCSOM and B2/191/P1/SWiM. 
The authors thank the anonymous referee for helpful remarks.

\newpage

\appendix

\section{RKD vs. SKD}\label{app:dist}

The regularised Kappa distribution (RKD) can be seen as a generalisation of the standard Kappa distribution (SKD)
and was introduced by \cite{Scherer-etal-2017} in the form
\begin{equation}\label{eq:rkd_general}
    f_{\mathrm{RKD}} = \frac{n}{\pi^{3/2} \,\Theta^3 \,\kappa^{3/2}\,\mathcal{U}_0} 
    \left(1 + \frac{v^2}{\kappa\,\Theta^2} \right)^{-(\kappa + 1)} 
    \exp\left(-\xi^2\frac{v^2}{\Theta^2}\right)\,,
\end{equation}
where $\mathcal{U}(a,b,x)$ is Kummer's function (see, e.g., \cite{Abramowitz-Stegun-1970}) 
and $\mathcal{U}_0 \equiv \mathcal{U}(3/2,3/2 - \kappa,\xi^2 \kappa)$. 
Furthermore, $n$ is the particle number density, $v$ is the individual particle speed, 
and $\kappa$ is a free parameter characterizing the high-energy 
tails of the distribution. 
The variable $\Theta$ is often termed thermal speed and can in principal be included either as a 
constant speed determined by observations, or as an analytical expression. 
In order to compare the results obtained for an RKD, an SKD and a Maxwellian, we choose the 
latter variant and use for $\Theta$ the Maxwellian thermal speed  
$\Theta = \sqrt{2\,k_\mathrm{B}\,T/m}$ with Boltzmann's 
constant $k_\mathrm{B}$, corresponding Maxwellian temperature $T$ and particle 
mass $m$.
The cutoff parameter $\xi$ has to fulfill the 
relation $\xi > \Theta/c$  \citep{Scherer-etal-2019} with vacuum speed of light $c$, but must be small 
enough to retain the main implication of the distribution.
By setting $\xi = 0$, the SKD is recovered and Eq.~\eqref{eq:rkd_general} becomes
\begin{equation}\label{eq:skd_general}
    f_{\mathrm{SKD}} = \frac{n}{\pi^{3/2} \,\Theta^3} 
    \frac{\Gamma(\kappa + 1)}{\kappa^{3/2}\,\Gamma(\kappa - 1/2)}
    \left(1 + \frac{v^2}{\kappa\,\Theta^2} \right)^{-(\kappa + 1)}\,,
\end{equation}
where $\Gamma (x)$ is the (complete) Gamma 
function of some argument x. If additionally $\kappa \to \infty$, the
Maxwellian distribution 
\begin{equation}\label{eq:maxwellian}
f_\mathrm{MD} = \frac{n}{\pi^{3/2}\,\Theta^3}\,\exp\left(- \frac{v^2}{\Theta^2} \right)
\end{equation}
is acquired.

\section{Useful formulas and integrals} \label{app:formulas}
In this appendix we present useful definitions, relations and general formulas for the calculations in Sec.~\ref{sec:tc}, which can be found in \cite{Abramowitz-Stegun-1970}, \cite{Oldham-etal-2000} and \cite{Scherer-etal-2020}. Kummer's function belongs to the confluent hypergeometric functions and can be represented in integral form as
\begin{equation}
    \mathcal{U}(a,b,x) = \frac{1}{\Gamma(a)} \int\limits_{0}^{\infty}\mathrm{d}t\,\exp(-x\,t)\,t^{\,a-1}\,(1+t)^{b-a-1} \quad\quad \mathrm{with}\,\, a > 0\,,\, x > 0.
\end{equation}
The case $a=0$ leads to $\mathcal{U}(0,b,x) = 1$, $\forall\, b,x \in \mathbb{R}$, while $x = 0$ not always yields finite solutions. However, if $b \leq 1$  and $x = 0$, then
\begin{equation}
    \mathcal{U}(a,b,0) = \frac{\Gamma(1-b)}{\Gamma(1 + a - b)}\,,
\end{equation}
which can be used to transform the RKD (Eq.~\ref{eq:rkd_general}) into the SKD (Eq.~\ref{eq:skd_general}). For a compact notation of the transport coefficients in Sec.~\ref{sec:tc} we further introduce the definition
\begin{equation}\label{eq:compact_kummer}
    \mathcal{U}_{[l,m]} \equiv \mathcal{U} \left(l, m - \kappa, \xi^2 \kappa \right)\,.
\end{equation}

The integrals in Sec.~\ref{sec:tc}, which calculate the $n-$th moment $M_n$ of the RKD, 
are of type 
\begin{equation}\label{eq:appendix_general_integral}
  M_n =   \Big \langle v^n\, (1 + A_\kappa\,v^2)^\eta \Big \rangle = 
    4\,\pi\,N_\kappa \int\limits_{0}^{\infty} \mathrm{d}v\, v^{\,n+2}\,(1 + A_\kappa v^2)^{- \zeta}\,\exp\left(-\xi^2\frac{v^2}{\Theta^2}\right)
\end{equation}
with $4\,\pi$ being the solution of the integrals over $\theta$ and $\phi$, $\eta$ being either -1 or 0, and $\zeta$ being either $\kappa+1$ or $\kappa+2$ in the integrals in Sec.~\ref{sec:tc}. General solutions
are given by \cite{Scherer-etal-2020} and read
\begin{equation}\label{eq:general_solutions}
    M_n = 2\,\pi\,\kappa^{\frac{3+n}{2}}\,\Theta^{3+n}\,\Gamma\left(\frac{3+n}{2}\right)\,\mathcal{U}\left(\frac{3+n}{2}, \frac{3+n}{2} - \zeta, \xi^2\,\kappa \right)\,.
\end{equation}
and thus $m = l$ or $m=l+1$ in Eq.~\eqref{eq:compact_kummer}.
Using Eqs.~\eqref{eq:appendix_general_integral} and \eqref{eq:general_solutions}, we obtain the following solutions for the integrals in Sec.~\ref{sec:tc}:
\begin{align}
    \Big \langle v^5\, (1 + A_\kappa\,v^2)^{-1} \Big \rangle &= 12\,\pi\,\Theta^8\,\kappa^4\,
    \mathcal{U}\left(4,3 - \kappa,\xi^2 \kappa \right) \label{eq:app_b_1} \\
    \Big \langle v^7\, (1 + A_\kappa\,v^2)^{-1} \Big \rangle &= 48\,\pi\,\Theta^{10}\,\kappa^5\,
    \mathcal{U}\left(5,4 - \kappa,\xi^2 \kappa \right) \label{eq_app_b_2} \\
    \Big \langle v^9\, (1 + A_\kappa\,v^2)^{-1} \Big \rangle &= 240\,\pi\,\Theta^{12}\,\kappa^6\,
    \mathcal{U}\left(6,5 - \kappa,\xi^2 \kappa \right) \label{eq:app_b_3} \\
    \Big \langle v^5 \Big \rangle &= 12\,\pi\,\Theta^8\,\kappa^4\,
    \mathcal{U}\left(4,4 - \kappa,\xi^2 \kappa \right) \label{eq:app_b_4} \\
    \Big \langle v^7 \Big \rangle &= 48\,\pi\,\Theta^{10}\,\kappa^5\,
    \mathcal{U}\left(5,5 - \kappa,\xi^2 \kappa \right) \label{eq:app_b_5} \\
    \Big \langle v^9 \Big \rangle &= 240\,\pi\,\Theta^{12}\,\kappa^6\,
    \mathcal{U}\left(6,6 - \kappa,\xi^2 \kappa \right) \label{eq:app_b_6}
\end{align}

For the derivation of the maximum values of the transport coefficients in the limit $\kappa \to 0$, the following relations are helpful and are taken from Tab.~A1 of \cite{Scherer-etal-2020}:
\begin{align}
    \mathcal{U}(a,b,\xi^2\,\kappa) &= \frac{\Gamma(c-1)}{\Gamma(a)\,(\xi^2\,\kappa)^{\,c-1}} + \frac{\Gamma(1-c)}{\Gamma(\kappa + 1)} \quad\quad\quad\,\,\,\, \mathrm{for}\,\, 1 < b < 2, \label{eq:case_5_scherer}\\
    \mathcal{U}(a,b,\xi^2\,\kappa) &= \frac{(d-2-(\kappa + 1)\,\xi^2\,\kappa)\,\Gamma(c-2)}{\Gamma(a)\,(\xi^2\,\kappa)^{c-1}} \quad\quad \mathrm{for}\,\, b > 2, \label{eq:case_7_scherer}
\end{align}
where $a = (3+n)/2$, $b = (3 + n - 2\,\kappa)/2$, $d= 1+a-b$, and moment $n \in \mathbb{N}_0$.
\section{Tabulated transport coefficients and their limits for \texorpdfstring{$\kappa \to 0$}{~}} \label{app:results}
The RKD is well defined for all $\kappa > 0$. While $\kappa = 0$ cannot be directly inserted into Kummer's function, the continuation $\kappa \to 0$ is still possible by using approximations. Thus, for the purpose of mathematical completeness, we derive the maximum values of the transport coefficients under consideration, which are obtained in the limit $\kappa \to 0$ (and $\xi > 0$). 
We begin by recognising that small but finite $\kappa$-values and finite $\xi$-values enable to make the following approximations for $\mathcal{U}_0$ and $\mathcal{U}_{[l,m]}$:
\begin{align}
   \mathcal{U}\left(\frac{3}{2},\frac{3}{2}-\kappa,\xi^2 \kappa\right)&\underset{\kappa \ll 1}{\approx}  \mathcal{U}\left(\frac{3}{2},\frac{3}{2},\xi^2 \kappa\right) \label{eq:u0_approx}\,,\\
    \mathcal{U}(l,m - \kappa, \xi^2 \kappa)&\underset{\kappa \ll 1}{\approx}  \mathcal{U}(l,m,\xi^2 \kappa)\,,\qquad \mathrm{for}\,\, m >2\,. \label{eq:u_approx}
\end{align}
We then may apply Eq.~\eqref{eq:case_5_scherer} to Eq.~\eqref{eq:u0_approx} and Eq.~\eqref{eq:case_7_scherer} to Eq.~\eqref{eq:u_approx} in order to obtain
\begin{align}
   \mathcal{U}\left(\frac{3}{2},\frac{3}{2},\xi^2 \kappa\right) &=  
   \frac{\Gamma\left(\frac{1}{2}\right)}{\Gamma\left(\frac{3}{2}\right)\, (\xi^{2} \kappa)^{1/2}} + \frac{\Gamma\left(-\frac{1}{2}\right)}{\Gamma(\kappa+1)}\,, \label{eq:case_5}\\
\mathcal{U}(l,m, \xi^2 \kappa) &=
\frac{\Big[m - 2 - (\kappa+1)\, \xi^2 \kappa\Big]\,\Gamma(m-2)}
{\Gamma(l)\,(\xi^2 \kappa)^{\,m-1}}\,. \label{eq:case_7}
\end{align}

We begin with the diffusion coefficient, where the approximation yields
\begin{equation}
    \frac{D}{D_\mathrm{M}} = \frac{\kappa^{5/2}\,
    \mathcal{U}_{[4,4]}}{\mathcal{U}_0}\ 
    \underset{\kappa \ll 1}{\approx}
    \frac{2 - \xi^2 \kappa}{6\,\xi^5}
    \frac{\Gamma\left(\frac{3}{2}\right)}{\Gamma\left(\frac{1}{2}\right) + \Gamma\left(-\frac{1}{2}\right)\,\Gamma\left(\frac{3}{2}\right)\,(\xi^2 \kappa)^{1/2}}\,.
\end{equation}
In the limit $\kappa \to 0$ we then obtain
\begin{equation}\label{eq:limit_D}
    \lim \limits_{\kappa \to 0} \frac{D}{D_\mathrm{M}} = \frac{1}{6\,\xi^5}\,.
\end{equation}
This result is in agreement with Fig.~\ref{fig:diffusion} for $\xi \in \{0.01,0.05,0.1\}$.

For both the mobility coefficient and the electric conductivity we obtain the same expression, which reads
\begin{equation}
    \begin{split}
        \frac{\mu}{\mu_\mathrm{M}} &= \frac{\sigma}{\sigma_\mathrm{M}} 
        = \frac{\kappa^{3/2}}{\mathcal{U}_0}
    \left[(\kappa+1)\,\mathcal{U}_{[4,3]} + \kappa\,\xi^2\,\mathcal{U}_{[4,4]} \right] \\
        &\underset{\kappa \ll 1}{\approx}
        \frac{3 - 2\,\xi^2 \kappa}{6\,\xi^3} 
    \frac{\Gamma\left(\frac{3}{2}\right)}{\Gamma\left(\frac{1}{2}\right) + \Gamma\left(-\frac{1}{2}\right)\,\Gamma\left(\frac{3}{2}\right)\,(\xi^2 \kappa)^{1/2}}\,,
    \end{split}
\end{equation}
which becomes in the limit $\kappa \to 0$
\begin{equation}
    \lim \limits_{\kappa \to 0} \frac{\mu}{\mu_\mathrm{M}} 
    = \lim \limits_{\kappa \to 0} \frac{\sigma}{\sigma_\mathrm{M}} =
    \frac{1}{4\,\xi^3}\,.
\end{equation}
We refer to Figs.~\ref{fig:mobility} and \ref{fig:sigma} for a comparison of the maximum values for $\xi \in \{0.01,0.05,0.1\}$.

\begin{deluxetable}{lllll}[t!]
\tablecaption{Overview of the transport coefficients. Displayed are the calculated transport coefficients (TC), i.e., diffusion coefficient $D$, mobility coefficient $\mu$, electric conductivity $\sigma$, thermoelectric coefficient $\alpha$ and thermal conductivity $\lambda$, based on different models, the Maxwellian (subscript M), the RKD (subscript RKD) and the SKD (subscript SKD). The last column shows the maximum values of the transport coefficients based on the $f_{\mathrm{r}\kappa}$ in the limit $\kappa \to 0$ (and $\xi > 0$).\label{tab:results}}
\tablewidth{700pt}
\tabletypesize{\scriptsize}
\tablehead{
\colhead{TC / Model} & \colhead{Maxwellian (M)} & \colhead{RKD} & \colhead{SKD} & \colhead{$\lim \limits_{\kappa \to 0}$}
} 
\startdata
Diffusion $D$ & $\frac{4\,\sqrt{2}\,(k_\mathrm{B}\,T)^{5/2}}{\pi^{3/2}\,m^{1/2}\,z\,e^4\,L^\mathrm{ei}\,n}$ & $D_\mathrm{M} \frac{\kappa^{5/2}\,
    \mathcal{U}_{[4,4]}}{\mathcal{U}_0}$ & $D_\mathrm{M} \frac{\Gamma(\kappa - 3)}{\Gamma(\kappa - 1/2)}\,\kappa^{5/2}$ & $\frac{D_\mathrm{M}}{6\,\xi^5}$\\
Mobility $\mu$ & $\frac{4\,\sqrt{2}\,(k_\mathrm{B}\,T)^{3/2}}{\pi^{3/2}\,m^{1/2}\,z\,e^3\,L^\mathrm{ei}\,n}$ & $\mu_\mathrm{M} \frac{\kappa^{3/2}}{\mathcal{U}_0}
    \left[(\kappa+1)\,\mathcal{U}_{[4,3]} + \kappa\,\xi^2\,\mathcal{U}_{[4,4]} \right]$ & $\mu_\mathrm{M} \frac{\Gamma\left(\kappa - 2\right)}{\Gamma\left(\kappa - 1/2\right)}\, \kappa^{3/2}$ & $ \frac{\mu_\mathrm{M}}{4\,\xi^3}$\\
Electric conductivity $\sigma$ & $\frac{4\,\sqrt{2}\,
    \left(k_\mathrm{B}\,T \right)^{3/2}}{m^{1/2}\,\pi^{3/2}\,z\,e^2\,L^\mathrm{ei}}$ &  $\sigma_\mathrm{M} \frac{\kappa^{3/2}}{\mathcal{U}_0}
    \left[(\kappa+1)\,\mathcal{U}_{[4,3]} + \kappa\,\xi^2\,\mathcal{U}_{[4,4]} \right]$ & $\sigma_\mathrm{M} \frac{\Gamma\left(\kappa - 2\right)}{\Gamma\left(\kappa - 1/2\right)}\, \kappa^{3/2}$ & $ \frac{\sigma_\mathrm{M}}{4\,\xi^3}$\\
Thermoelectric coefficient $\alpha$ & $-\frac{5}{2} \frac{k_\mathrm{B}}{e}$ & $\alpha_\mathrm{M} \frac{\kappa\,\left[4\,(\kappa+1)\,\mathcal{U}_{[5,4]} -\frac{3}{2}\,
    \mathcal{U}_{[4,4]}
    + 4\,\kappa\,\xi^2\,
    \mathcal{U}_{[5,5]}\right]}{5/2\,\left[(\kappa+1)\,\mathcal{U}_{[4,3]} + \kappa\,\xi^2\,\mathcal{U}_{[4,4]}\right]}$ & $\alpha_\mathrm{M}\frac{\kappa}{\kappa - 3}$ & $\frac{2\,\alpha_\mathrm{M}}{3\,\xi^2}$\\
Thermal conductivity $\lambda$ & $\frac{16\,\sqrt{2}\,k_\mathrm{B}\,\left(k_\mathrm{B}\,T \right)^{5/2}}
    {m^{1/2}\,\pi^{3/2}\,z\,e^4\,L^\mathrm{ei}}$ & $\lambda_\mathrm{M} \frac{\kappa^{7/2}}{\mathcal{U}_0} 
    \Biggl[5(\kappa + 1)\,\mathcal{U}_{[6,5]}
    + 5\,\xi^2\,\kappa\,\mathcal{U}_{[6,6]}
    -\frac{3}{2}\,\mathcal{U}_{[5,5]}$ 
    & $\lambda_\mathrm{M} \frac{\Gamma(\kappa - 4)}{\Gamma\left(\kappa - 1/2\right)}\,
    \frac{\kappa^{7/2}\,(\kappa - 1/2)}{(\kappa - 3)}$ & $\frac{23\,\lambda_\mathrm{M}}{144\,\xi^7}$\\
    & & $- \left[(\kappa+1)\, \mathcal{U}_{[5,4]}
    + \xi^2\,\kappa\,\mathcal{U}_{[5,5]}\right] 
      \frac{4\,(\kappa+1)\,\mathcal{U}_{[5,4]} -\frac{3}{2}\,
    \mathcal{U}_{[4,4]}
    + 4\,\kappa\,\xi^2\,
    \mathcal{U}_{[5,5]}}{(\kappa+1)\,\mathcal{U}_{[4,3]} + \kappa\,\xi^2\,\mathcal{U}_{[4,4]}}
     \Biggr]$ & & \\
\enddata
\end{deluxetable}

The thermoelectric coefficient can be written for $\kappa \ll 1$ as
\begin{equation}\label{eq:alpha_kll1}
    \begin{split}
        \frac{\alpha}{\alpha_\mathrm{M}} &= \frac{\kappa}{5}\, \frac{8\,(\kappa+1)\,\mathcal{U}_{[5,4]} -3\,
    \mathcal{U}_{[4,4]}
    + 8\,\xi^2 \kappa\,
    \mathcal{U}_{[5,5]}}{(\kappa+1)\,\mathcal{U}_{[4,3]} + \xi^2\,\kappa\,\mathcal{U}_{[4,4]}} \\
    &\underset{\kappa \ll 1}{\approx}
    \frac{6}{3 - 2\,\xi^2 \kappa} 
    \times \left(\frac{2 - \xi^2 \kappa}{15\,\xi^2} 
    - \frac{2 - \xi^2 \kappa}{10\,\xi^2} 
    + \frac{6 - 2\,\xi^2 \kappa}{15\,\xi^2} \right)\,,
    \end{split}
\end{equation}
which yields for limit $\kappa \to 0$
\begin{equation}
    \lim \limits_{\kappa \to 0} \frac{\alpha}{\alpha_\mathrm{M}} = \frac{2}{3\,\xi^2}.
\end{equation}
This result is in agreement with the corresponding plot in
Fig.~\ref{fig:alpha} for $\xi \in \{0.01,0.05,0.1\}$.

Finally, the thermal conductivity can be approximated as
\begin{equation}\label{eq:max_lambda}
    \begin{split}
        \frac{\lambda}{\lambda_\mathrm{M}} &= \frac{\kappa^{7/2}}{\mathcal{U}_0} 
    \Biggl[5(\kappa + 1)\,\mathcal{U}_{[6,5]} 
    + 5\,\xi^2\,\kappa\,\mathcal{U}_{[6,6]}
    -\frac{3}{2}\,\mathcal{U}_{[5,5]} \\
    &- \left[(\kappa+1)\, \mathcal{U}_{[5,4]}
    + \xi^2\,\kappa\,\mathcal{U}_{[5,5]}\right] 
     \frac{4\,(\kappa+1)\,\mathcal{U}_{[5,4]} -\frac{3}{2}\,
    \mathcal{U}_{[4,4]}
    + 4\,\kappa\,\xi^2\,
    \mathcal{U}_{[5,5]}}{(\kappa+1)\,\mathcal{U}_{[4,3]} + \kappa\,\xi^2\,\mathcal{U}_{[4,4]}}
     \Biggr] \\
      &\underset{\kappa \ll 1}{\approx}
      \frac{\Gamma\left(\frac{3}{2}\right)}{\Gamma\left(\frac{1}{2}\right) + \Gamma\left(-\frac{1}{2}\right)\,\Gamma\left(\frac{3}{2}\right)\,(\xi^2 \kappa)^{1/2}} 
       \Bigg\{\frac{3 - \xi^2 \kappa}{12\,\xi^7} 
      + \frac{4 - \xi^2 \kappa}{4\,\xi^7} \\
      &- \frac{3 - \xi^2 \kappa}{8\,\xi^7} 
      - \frac{8 - 3\,\xi^2 \kappa}{24\,\xi^7}
       \frac{6}{3 - 2\,\xi^2 \kappa}
      \frac{10 - 3\,\xi^2 \kappa}{12}
      \Bigg\}\,.
    \end{split}
\end{equation}
In the limit $\kappa \to 0$, Eq.~\eqref{eq:max_lambda} turns into
\begin{equation}
    \lim \limits_{\kappa \to 0} \frac{\lambda}{\lambda_\mathrm{M}} = \frac{23}{144\,\xi^7}.
\end{equation}
The result can be compared to Fig.~\ref{fig:lambda} for $\xi \in \{0.01,0.05,0.1\}$.

Many terms in the transport coefficients are of type $\kappa^{\,n/2}\,\mathcal{U}_{[l,m]}/\mathcal{U}_0$, that is, 
\begin{align}
\chi \equiv \kappa^{\,n/2} \frac{\mathcal{U}\left(\frac{n+3}{2},\frac{n+5}{2}-\zeta,\kappa\,\xi^2\right)}{\mathcal{U}\left(\frac{3}{2},\frac{3}{2},\kappa\,\xi^2\right)}\,,
\end{align}
where $l = (n+3)/2$ and $m = (n+5)/2 - \zeta$, for which the limit $\kappa \to 0$ can be estimated as follows. Considering $\kappa \ll 1$, if $\zeta = \kappa + 1 \approx 1$, then
\begin{align}
\chi &\approx \kappa^{n/2} \frac{\mathcal{U}\left(\frac{n+3}{2},\frac{n+3}{2},\kappa\,\xi^2\right)}{\mathcal{U}\left(\frac{3}{2},\frac{3}{2},\kappa\,\xi^2\right)} \\
&= \kappa^{n/2} \frac{\left(\frac{n-1}{2} - \kappa\,\xi^2 \right)\,\Gamma\left(\frac{n-1}{2} \right)}{\Gamma\left(\frac{n+3}{2}\right)\,\left(\kappa\,\xi^2\right)^{(n+1)/2}}\,
\frac{\Gamma\left(\frac{3}{2}\right)\,\left(\kappa\,\xi^2\right)^{1/2}}{\Gamma\left(\frac{1}{2}\right) + \Gamma\left(-\frac{1}{2}\right)\,\Gamma\left(\frac{3}{2}\right)\,\left(\kappa\,\xi^2\right)^{1/2}} \\
&= \frac{\left(\frac{n-1}{2} - \kappa\,\xi^2 \right)\,\Gamma\left(\frac{n-1}{2} \right)}{\Gamma\left(\frac{n+3}{2}\right)\,\xi^n}\,
\frac{\Gamma\left(\frac{3}{2}\right)}{\Gamma\left(\frac{1}{2}\right) + \Gamma\left(-\frac{1}{2}\right)\,\Gamma\left(\frac{3}{2}\right)\,\left(\kappa\,\xi^2\right)^{1/2}}\,,
\end{align}
where we again made use of Eqs.~\eqref{eq:case_5_scherer} and \eqref{eq:case_7_scherer}.
In the limit $\kappa \to 0$ we then obtain
\begin{align}
\lim_{\kappa \to 0} \chi &= 
\frac{\left(\frac{n-1}{2} \right)\,\Gamma\left(\frac{n-1}{2} \right)\,\Gamma\left(\frac{3}{2} \right)}{\Gamma\left(\frac{n+3}{2}\right)\,\xi^n\,\Gamma\left(\frac{1}{2} \right)} \\
&= \frac{1}{(n+1)\,\xi^n}\,.
\end{align}
The case $\zeta = \kappa + 2 \approx 2$ yields
\begin{align}
\chi &\approx \kappa^{n/2} \frac{\mathcal{U}\left(\frac{n+3}{2},\frac{n+1}{2},\kappa\,\xi^2\right)}{\mathcal{U}\left(\frac{3}{2},\frac{3}{2},\kappa\,\xi^2\right)} \\
&= \kappa^{n/2} \frac{\left(\frac{n-3}{2} - \kappa\,\xi^2 \right)\,\Gamma\left(\frac{n-3}{2} \right)}{\Gamma\left(\frac{n+3}{2}\right)\,\left(\kappa\,\xi^2\right)^{(n-1)/2}}\,
\frac{\Gamma\left(\frac{3}{2}\right)\,\left(\kappa\,\xi^2\right)^{1/2}}{\Gamma\left(\frac{1}{2}\right) + \Gamma\left(-\frac{1}{2}\right)\,\Gamma\left(\frac{3}{2}\right)\,\left(\kappa\,\xi^2\right)^{1/2}} \\
&= \kappa \frac{\left(\frac{n-3}{2} - \kappa\,\xi^2 \right)\,\Gamma\left(\frac{n-3}{2} \right)}{\Gamma\left(\frac{n+3}{2}\right)\,\xi^{2-n}}\,
\frac{\Gamma\left(\frac{3}{2}\right)}{\Gamma\left(\frac{1}{2}\right) + \Gamma\left(-\frac{1}{2}\right)\,\Gamma\left(\frac{3}{2}\right)\,\left(\kappa\,\xi^2\right)^{1/2}}\,,
\end{align}
which in the limit $\kappa \to 0$ becomes
\begin{align}
\lim_{\kappa \to 0} \chi &= \kappa
\frac{\left(\frac{n-3}{2} \right)\,\Gamma\left(\frac{n-3}{2} \right)\,\Gamma\left(\frac{3}{2} \right)}{\Gamma\left(\frac{n+3}{2}\right)\,\xi^{2-n}\,\Gamma\left(\frac{1}{2} \right)} \\
&= 0\,.
\end{align}

Table~\ref{tab:results} summarises the results of this manuscript and contains the transport coefficients based on the Maxwellian distribution (subscript M), on the regularised Kappa distribution (RKD) and on the Standard Kappa distribution (SKD). The last column shows the maximum values of the transport coefficients based on the RKD, obtained in the limit $\kappa \to 0$.

\bibliography{rkd}{}

\begin{thebibliography}{}
\expandafter\ifx\csname natexlab\endcsname\relax\def\natexlab#1{#1}\fi
\providecommand{\url}[1]{\href{#1}{#1}}
\providecommand{\dodoi}[1]{doi:~\href{http://doi.org/#1}{\nolinkurl{#1}}}
\providecommand{\doeprint}[1]{\href{http://ascl.net/#1}{\nolinkurl{http://ascl.net/#1}}}
\providecommand{\doarXiv}[1]{\href{https://arxiv.org/abs/#1}{\nolinkurl{https://arxiv.org/abs/#1}}}

\bibitem[{{Abramowitz} \& {Stegun}(1970)}]{Abramowitz-Stegun-1970}
{Abramowitz}, M., \& {Stegun}, I. 1970, {Handbook of mathematical functions}
  (New York: Dover Publications)

\bibitem[{{Balescu}(1988)}]{Balescu-1988}
{Balescu}, R. 1988, {Transport processes in plasmas - 1. Classical transport
  theory} (Amsterdam: North Holland Publishing Company)

\bibitem[{{Bhatnagar} {et~al.}(1954){Bhatnagar}, {Gross}, \&
  {Krook}}]{Bhatnagar-etal-1954}
{Bhatnagar}, P.~L., {Gross}, E.~P., \& {Krook}, M. 1954, Phys. Rev., 94, 511

\bibitem[{{Bian} {et~al.}(2014){Bian}, {Emslie}, {Stackhouse1}, \&
  {Kontar}}]{Bian-etal-2014}
{Bian}, N.~H., {Emslie}, A.~G., {Stackhouse1}, D.~J., \& {Kontar}, E.~P. 2014,
  ApJ, 796, 142

\bibitem[{{Bourouaine} {et~al.}(2011){Bourouaine}, {Marsch}, \&
  {Neubauer}}]{Bourouaine-etal-2011}
{Bourouaine}, S., {Marsch}, E., \& {Neubauer}, F.~M. 2011, A\&A, 536, A39

\bibitem[{{Boyd} \& {Sanderson}(2003)}]{Boyd-Sanderson-2003}
{Boyd}, T.~J.~M., \& {Sanderson}, J.~J. 2003, {The Physics of Plasmas}
  (Cambridge: Cambridge University Press)

\bibitem[{{Braginskii}(1965)}]{Braginskii-1965}
{Braginskii}, S.~I. 1965, Rev. Plasma Phys., 1, 205

\bibitem[{{Dum}(1990)}]{Dum-1990}
{Dum}, C.~T. 1990, in Physical Processes in Hot Cosmic Plasmas, ed.
  W.~{Brinkmann}, A.~C. {Fabian}, \& F.~{Giovanelli} (Dordrecht/Boston/London:
  Kluwer Academic Publishers), 157--180

\bibitem[{{Ebne Abbasi} {et~al.}(2017){Ebne Abbasi}, {Esfandyari-Kalejahi}, \&
  {Khaledi}}]{Abbasi-etal-2017}
{Ebne Abbasi}, Z., {Esfandyari-Kalejahi}, A., \& {Khaledi}, P. 2017, Astrophys.
  Space Sci., 362, 103

\bibitem[{{Echim} {et~al.}(2011){Echim}, {Lemaire}, \&
  {Lie-Svendsen}}]{Echim-etal-2011}
{Echim}, M.~M., {Lemaire}, J., \& {Lie-Svendsen}, {\O}. 2011, Surv. Geophys.,
  32, 1

\bibitem[{{Goedbloed} {et~al.}(2019){Goedbloed}, {Keppens}, \&
  {Poedts}}]{Goedbloed-etal-2019}
{Goedbloed}, H., {Keppens}, R., \& {Poedts}, S. 2019, {Magnetohydrodynamics of
  Laboratory and Astrophysical Plasmas} (Cambridge: Cambridge University Press)

\bibitem[{{Guo} \& {Du}(2019)}]{Guo-Du-2019}
{Guo}, R., \& {Du}, J. 2019, Physica A, 523, 156

\bibitem[{{Helander} \& {Sigmar}(2005)}]{Helander-Sigmar-2005}
{Helander}, P., \& {Sigmar}, D.~J. 2005, {Collisional Transport in Magnetized
  Plasmas} (Cambridge: Cambridge University Press)

\bibitem[{{Husidic} {et~al.}(2021){Husidic}, {Lazar}, {Fichtner}, \&
  {Poedts}}]{Husidic-etal-2021}
{Husidic}, E., {Lazar}, M., {Fichtner}, H., \& {Poedts}, S. 2021, A\&A, 654,
  A99

\bibitem[{{Landi}(2007)}]{Landi-2007}
{Landi}, E. 2007, ApJ, 663, 1363

\bibitem[{{Landi} \& {Cranmer}(2009)}]{Landi-Cranmer-2009}
{Landi}, E., \& {Cranmer}, S.~R. 2009, ApJ, 691, 794

\bibitem[{{Landi} {et~al.}(2010){Landi}, {Pantellini}, \&
  {Matteini}}]{Landi-etal-2010}
{Landi}, S., {Pantellini}, F., \& {Matteini}, L. 2010, AIP Conf. Proc., 1216,
  218

\bibitem[{{Landi} {et~al.}(2012){Landi}, {Pantellini}, \&
  {Matteini}}]{Landi-etal-2012}
---. 2012, ApJ, 760, 143

\bibitem[{{Lazar} \& {Fichtner}(2021)}]{Lazar-Fichtner-2021}
{Lazar}, M., \& {Fichtner}, H., eds. 2021, Kappa Distributions: From
  Observational Evidences via Controversial Predictions to a Consistent Theory
  of Non-equilibrium Plasmas (Springer)

\bibitem[{{Lazar} {et~al.}(2016){Lazar}, {Fichtner}, \&
  {Yoon}}]{Lazar-etal-2016}
{Lazar}, M., {Fichtner}, H., \& {Yoon}, P.~H. 2016, A\&A, 589, A39

\bibitem[{{Lazar} {et~al.}(2015){Lazar}, {Poedts}, \&
  {Fichtner}}]{Lazar-etal-2015}
{Lazar}, M., {Poedts}, S., \& {Fichtner}, H. 2015, A\&A, 582, A124

\bibitem[{{Lazar} {et~al.}(2020){Lazar}, {Scherer}, {Fichtner}, \&
  {Pierrard}}]{Lazar-etal-2020}
{Lazar}, M., {Scherer}, K., {Fichtner}, H., \& {Pierrard}, V. 2020, A\&A, 634,
  A20

\bibitem[{{Maksimovic} {et~al.}(2005){Maksimovic}, {Zouganelis}, {Chaufray},
  {Issautier}, {Scime}, {Littleton}, {Marsch}, {McComas}, {Salem}, {Lin}, \&
  {Elliott}}]{Maksimovic-etal-2005}
{Maksimovic}, M., {Zouganelis}, I., {Chaufray}, J.~Y., {et~al.} 2005, J.
  Geophys. Res., 110, A09104

\bibitem[{{Marsch}(1994)}]{Marsch-1994}
{Marsch}, E. 1994, Adv. Space Res., 14, 103

\bibitem[{{Marsch}(2006)}]{Marsch-2006}
---. 2006, Liv. Rev. Solar Phys., 3, 1

\bibitem[{{Olbert}(1968)}]{Olbert-1968}
{Olbert}, S. 1968, Phys. Magnetos., 10, 641

\bibitem[{{Oldham} {et~al.}(2009){Oldham}, {Myland}, \&
  {Spanier}}]{Oldham-etal-2000}
{Oldham}, K., {Myland}, J., \& {Spanier}, J. 2009, {An Atlas of Functions} (New
  York: Springer Science$+$Business Media)

\bibitem[{{Pierrard} \& {Lazar}(2010)}]{Pierrard-Lazar-2010}
{Pierrard}, V., \& {Lazar}, M. 2010, Sol. Phys., 267, 153

\bibitem[{{Salem} {et~al.}(2003){Salem}, {Hubert}, {Lacombe}, {Bale},
  {Mangeney}, {Larson}, \& {Lin}}]{Salem-etal-2003}
{Salem}, C., {Hubert}, D., {Lacombe}, C., {et~al.} 2003, ApJ, 585, 1147

\bibitem[{{Scherer} {et~al.}(2017){Scherer}, {Fichtner}, \&
  {Lazar}}]{Scherer-etal-2017}
{Scherer}, K., {Fichtner}, H., \& {Lazar}, M. 2017, Europhys. Lett., 120, 50002

\bibitem[{{Scherer} {et~al.}(2019){Scherer}, {Fichtner}, \&
  {Lazar}}]{Scherer-etal-2019}
{Scherer}, K., {Fichtner}, H.~{Fahr}, H.~J., \& {Lazar}, M. 2019, ApJ, 881, 93

\bibitem[{{Scherer} {et~al.}(2020){Scherer}, {Husidic}, {Lazar}, \&
  {Fichtner}}]{Scherer-etal-2020}
{Scherer}, K., {Husidic}, E., {Lazar}, M., \& {Fichtner}, H. 2020, MNRAS, 497,
  1738

\bibitem[{{Spatschek}(1990)}]{Spatschek-1990}
{Spatschek}, K.~H. 1990, {Theoretische Plasmaphysik - Eine Einf\"{u}hrung}
  (Stuttgart: B.~G. Teubner Verlag)

\bibitem[{{Vasyli\={u}nas}(1968)}]{Vasyliunas-1968}
{Vasyli\={u}nas}, V.~M. 1968, J. Geophys. Res.: Space Phys., 73, 2839

\bibitem[{{Verscharen} {et~al.}(2019){Verscharen}, {Klein}, \&
  {Maruca}}]{Verscharen-etal-2019}
{Verscharen}, D., {Klein}, K.~G., \& {Maruca}, B.~A. 2019, Living Rev. Sol.
  Phys., 16, 5

\bibitem[{{Vocks} \& {Mann}(2003)}]{Vocks-Mann-2003}
{Vocks}, C., \& {Mann}, G. 2003, ApJ, 593, 1134

\bibitem[{{Vocks} {et~al.}(2008){Vocks}, {Mann}, \&
  {Rausche}}]{Vocks-etal-2008}
{Vocks}, C., {Mann}, G., \& {Rausche}, G. 2008, A\&A, 480, 527

\bibitem[{{Wang} \& {Du}(2017)}]{Wang-Du-2017}
{Wang}, L., \& {Du}, J. 2017, Phys. Plasmas, 24, 102305

\bibitem[{{Yoon}(2011)}]{Yoon-2011}
{Yoon}, P.~H. 2011, Phys. Plasmas, 18, 122303

\bibitem[{{Yoon}(2014)}]{Yoon-2014}
---. 2014, J. Geophys. Res. Space Phys., 119, 7074

\bibitem[{{Zouganelis} {et~al.}(2005){Zouganelis}, {Meyer-Vernet}, {Landl},
  {Maksimovic}, \& {Pantellini}}]{Zouganelis-etal-2005}
{Zouganelis}, I., {Meyer-Vernet}, N., {Landl}, S., {Maksimovic}, M., \&
  {Pantellini}, F. 2005, ApJ, 626, L117

\end{thebibliography}
\bibliographystyle{aasjournal}

\end{document}